%% file: SUS-10-011_temp.tex
\pdfoutput=1

\documentclass[11pt,twoside,a4paper,cmspaper,final,collab]{cms-tdr}
\def\svnVersion{61640:61668}\def\cmsCernNo{2011-067}\def\cmsCernDate{\today}\def\cmsMessage{Submitted to the Journal of High Energy Physics}

\begin{document}\cmsNoteHeader{SUS-10-011}

\hyphenation{had-ron-i-za-tion}
\hyphenation{cal-or-i-me-ter}
\hyphenation{de-vices}
\RCS$Revision: 61668 $
\RCS$HeadURL: svn+ssh://alverson@svn.cern.ch/reps/tdr2/papers/SUS-10-011/trunk/SUS-10-011.tex $
\RCS$Id: SUS-10-011.tex 61668 2011-06-16 16:58:06Z alverson $
\newcommand{\Ht}{\ensuremath{H_{\mathrm{T}}}\xspace}
\newcommand{\deltaHt}{\ensuremath{\Delta H_{\mathrm{T}}}\xspace}
\newcommand{\gthreejet}{\mbox{$\geq3$~jet}\xspace}
\newcommand{\dphistar}{\mbox{$\Delta\phi^{*}$}\xspace}

\newcommand{\sigmaIetaIeta}{\ensuremath{\sigma_{\mathrm{i}\eta\mathrm{i}\eta}}\xspace}
\newcommand{\acceff}{\mbox{$A\times\epsilon$}\xspace}
\newcommand{\beff}{\mbox{$\epsilon_{b}$}\xspace}
\newcommand{\ptrel}{\ensuremath{p_{\mathrm{T}}^{\mathrm{rel}}}\xspace}
\newcommand{\Nobs}{\ensuremath{N^{\mathrm{obs}}_{95}}\xspace}
\newcommand{\sigmaobs}{\ensuremath{\sigma^{\mathrm{obs}}_{95}}\xspace}
\newcommand{\alphat}{\ensuremath{\alpha_{\mathrm{T}}}\xspace}
\newcommand{\mht}{\ensuremath{H\!\!\!\!/_{\mathrm{T}}}\xspace}
\newcommand{\gt}{\mbox{$>$}\xspace}
\newcommand{\lt}{\mbox{$<$}\xspace}
\newcommand{\Zmumu}{\mbox{${\rm Z} \rightarrow \mu^{+}\mu^{-}$}\xspace}
\newcommand{\Znunu}{\mbox{${\rm Z} \rightarrow \nu\overline{\nu}$}\xspace}
\newcommand{\FaT}{\ensuremath{\mathrm{F}(\alpha_{\mathrm{T}}>0.55)}\xspace}

\cmsNoteHeader{SUS-10-011}
\title{Search for Supersymmetry in Events with b Jets and
Missing Transverse Momentum at the LHC}

\address[cern]{CERN}
\author[cern]{The CMS Collaboration}

\date{\today}

\abstract{
 A search for supersymmetry is presented using a sample of events with
 b jets and missing transverse momentum.
The search uses a data sample of proton-proton collisions at a
centre-of-mass energy of 7~TeV, corresponding to an integrated
luminosity of 35~pb$^{-1}$, collected with the CMS detector.
A total of $0.33^{~+0.43}_{~-0.33}~(\mathrm{stat.}) \pm
  0.13~(\mathrm{syst.})$ events is predicted,
using control samples in the data, to arise from standard model
processes, and one event is observed in the data.
Upper limits are set at the $95\%$ confidence level on the cross
sections of benchmark supersymmetric models.
}

\hypersetup{%
pdfauthor={CMS Collaboration},%
pdftitle={Search for Supersymmetry in Events with b Jets and
Missing Transverse Momentum at the LHC},%
pdfsubject={CMS}, %
pdfkeywords={CMS, physics, SUSY}}

\maketitle 

\section{Introduction}
\label{sec:intro}

Supersymmetry (SUSY)~\cite{SUSY1,SUSY2,SUSY3,SUSY4,SUSY5} is an
extension of the standard model
(SM) of particle physics, which can solve the ``hierarchy
problem''~\cite{hierarchy1,hierarchy2} and provide a candidate for
cold dark matter~\cite{dark_matter}.
For a large class of supersymmetric parameter sets, squarks ($\tilde{\mathrm{q}}$),
the SUSY partners of quarks, are relatively light.  In this case,
significant event yields at the Large Hadron Collider (LHC) can result
from strong production of squarks,
which subsequently decay giving a weakly interacting lightest
supersymmetric particle (LSP).
If bottom and top squarks,
which can decay to b quarks, are relatively light, there may be an
abundance of events with one or more b-quark jets and momentum
imbalance transverse to the beam line due to the undetectable LSPs.

This Letter describes a search for events with two or more hadronic
jets, at least one of which must be b tagged~\cite{BTV10-001}, and
significant transverse momentum imbalance.  It
extends a similar search without a b-tag
requirement~\cite{main_RA1_note}.
The momentum imbalance is characterized~\cite{randall} by the ratio of
the \pt of the second-highest-\pt jet and the invariant mass formed
from the two highest-\pt jets.  This ratio can be estimated by
$\alphat=\frac{1}{2}\frac{\Ht-\deltaHt}{\sqrt{\Ht^2-\mht^2}}$,
where
$\mht = |\sum_{i}\vec{p}_{\mathrm{T}}^{{\rm ~jet}_i}|$,
$\Ht = \sum_{i}p_{\mathrm{T}}^{{\rm ~jet}_i}$, and
$p_{\mathrm{T}}^{{\rm ~jet}_i}$ is the momentum
transverse to the beam line for jet $i$ in an event.
The jets in an event are grouped into two pseudo-jets and \deltaHt is
the minimal value of
$|p_{\mathrm{T}}^{\mathrm{pseudojet}1}-p_{\mathrm{T}}^{\mathrm{pseudojet}2}|$
over all combinations; this approach optimizes rejection of
backgrounds with apparent \mht from instrumental effects and other sources.

The main backgrounds are
due to standard model multijet production (hereafter denoted ``QCD
background''),
electroweak W and Z boson production (EWK), and
top quark pair production (\ttbar).
Owing to low average \mht, the QCD background is effectively rejected
by a requirement on \alphat.  The b-tag requirement further suppresses
the QCD and EWK backgrounds.

The results of the search are characterized in terms of the
mSUGRA/CMSSM~\cite{mSUGRA,CMSSM} scenario of SUSY.
These models are described by four
parameters and one sign: the universal scalar and gaugino mass
parameters, $m_{0}$ and $m_{1/2}$, respectively; the universal
trilinear coupling, $A_{0}$; the ratio of the two Higgs doublet vacuum
expectation values, $\tan{\beta}$; and the sign of the Higgs mixing
parameter,  sign($\mu$).  Three signal points are considered as
benchmarks: LM0, LM1, both discussed
in Ref.~\cite{main_RA1_note}, and LMB (corresponding to
$m_{0}=400$~GeV, $m_{1/2}=200$~GeV, $A_{0}=0$~GeV, $\tan{\beta}=50$, and sign$(\mu)>0$), chosen to be near  the edge of
sensitivity of this search in mSUGRA/CMSSM parameter space.

The analysis presented here uses a data sample of proton-proton
collisions at 7~TeV, corresponding to an
integrated luminosity of $35~\mathrm{pb}^{-1}$, collected with the
Compact Muon Solenoid (CMS) detector, at the LHC.
The main components of CMS are a silicon pixel and strip tracker, the crystal
electromagnetic calorimeter, and the brass/scintillator hadron calorimeter, all
placed in a 3.8 T axial magnetic field, complemented by gas-ionization detectors
embedded in the steel return yoke, to measure muons.
A detailed description of the detector and its performance
can be found in Ref.~\cite{ref:CMS}.
In the cylindrical coordinate system of CMS, $\phi$ is the  azimuthal
angle and the pseudo-rapidity
($\eta$) is defined as $\eta = -\ln{[\tan{(\theta/2)}]}$, where $\theta$ is
the polar angle with respect to the counterclockwise beam direction.

\section{Event Selection}
\label{sec:event_selection}

The event selection requirements are mostly identical to those in
Ref.~\cite{main_RA1_note}.
Events in the search sample are collected with
triggers based on \Ht computed from jets
reconstructed at trigger level.  A muon-enriched control sample is
collected with triggers requiring a muon.  Events must have a
good reconstructed pp collision vertex~\cite{primary_vertex}.
Jets are reconstructed as clusters of energy in the
calorimeters by the anti-$\mathrm{k}_{\mathrm{T}}$
algorithm~\cite{anti-kT}
with a distance parameter of 0.5,
and are required to have
energy transverse to the beam, \et, in excess
of 50 GeV and $|\eta|$ less than 3.

To perform a fully hadronic final state search and to reduce the
backgrounds, events with an isolated lepton (electron or muon) or photon are
vetoed, and events consistent with having apparent
\mht~\cite{main_RA1_note} are rejected.
Selected events are required to have at least two jets, both with
\et\gt100~GeV, $|\eta|$~\lt2.5 for the
highest-\et jet, \HT\gt 350~GeV, at least one jet tagged as
originating from a b quark, and
\alphat\gt0.55.

Jets are b tagged using a discriminator based on the impact parameter
significance of tracks in a jet (Track Counting High Purity
discriminator, TCHP~\cite{BTV10-001}), with a ``tight'' selection
(TCHP\gt3.41) designed to have a light-flavour contamination of less than 0.1\%.
Looser b-tagging selections are used to produce various control samples.  An event
is said to be anti-tagged if it contains no jets with a loose b tag
(TCHP~\gt1.19).
To remain within the acceptance of
the pixel tracker, only jets with a central axis of $|\eta|$~\lt2.4 are
considered for b tagging.

\section{Background Estimation}
\label{sec:background}

The backgrounds for this search can be categorized into three main
groups: namely QCD, EWK, and \ttbar.  The contamination from \ttbar is
mainly in the tau decay mode.
The vast majority of events from the QCD background do not feature
large transverse momentum imbalance
and are therefore rejected by the \alphat\gt0.55 requirement.
The EWK backgrounds consist of W and Z boson production,
with genuine missing energy due to decay neutrinos.
The requirement of at least one
b jet greatly reduces the EWK and QCD backgrounds.
The dominant background for the analysis arises from \ttbar production,
in which b jets and genuine missing energy due to neutrinos can arise
from the top quark decay chains.

A procedure based on control data samples, described in
Section~\ref{sec:alphaT-vs-HT}, is employed to estimate
all backgrounds simultaneously.  In this method, the fraction of all
events with \alphat\gt0.55, denoted \FaT, is measured in a lower-\HT control
region and applied in the signal region.

The \Znunu and \ttbar background yields are cross-checked separately,
as discussed
in Section~\ref{sec:cross_checks}.  The
\ttbar cross-check uses muons to emulate the hadronic decays of
taus.  The cross-check of \Znunu utilizes \Zmumu events for which
\alphat is determined after excluding the muons.

\subsection{\texorpdfstring{Background Prediction Using \alphat vs \HT Extrapolation}{Background Prediction Using alphat vs HT Extrapolation}}
\label{sec:alphaT-vs-HT}

In SM simulation studies~\cite{main_RA1_note}, \FaT has no \HT
dependence in events with large genuine missing transverse energy,
i.e., the \ttbar and EWK backgrounds.
In the QCD background, however, \FaT is expected to be a decreasing
function of \HT because of
the \HT dependence of the factors contributing to apparent \mht, such as
jet energy resolution and jet \et threshold effects.

In data control samples, \FaT is consistent with having no
\HT dependence, which indicates that the \ttbar and EWK backgrounds
dominate.
The larger anti-tagged data sample is also
consistent with having no \HT dependence.
Because a tight b-tag
requirement further suppresses the QCD background,
the tight tagged data sample is expected to have a negligible QCD
contribution and therefore \FaT independent of \HT.

The total background is estimated by measuring
$\FaT =1.48^{+1.93}_{-1.48} \times 10^{-5}$ in a control region with 250~\lt\HT\lt350~GeV and multiplying this
fraction by the number of events in the signal region before the
\alphat\gt0.55 requirement.
In data, this procedure yields a prediction of
$0.33^{~+0.43}_{~-0.33}~(\mathrm{stat.}) \pm 0.13~(\mathrm{syst.})$ events.   The
statistical uncertainty is dominated by the
presence of one event with \alphat\gt0.55 in the control sample.
The systematic uncertainty on the prediction is given by the difference in
\FaT measured in the tight and loose tagged control samples.
Table~\ref{table:bkg_pred} lists this background prediction, the
observation in data, and the expected contribution
of SUSY signal for points LM0, LM1, and LMB.

\begin{table}[tb]
\caption[Fit Parameters] {\label{table:bkg_pred}
Predicted and observed numbers of events for 35~\pbinv.  The prediction comes
from the \alphat vs \HT extrapolation described in
Section~\ref{sec:alphaT-vs-HT}.
}
\renewcommand{\arraystretch}{1.3}

\begin{center}
\begin{tabular} { | l | c c c c c |}
\hline

N-jets & Background Prediction & Data & LM0 & LM1 & LMB \\
\hline

$\geq2$  & $0.33^{~+0.43}_{~-0.33}~(\mathrm{stat.}) \pm
0.13~(\mathrm{syst.})$    & 1  & 14 & 2 & 5 \\

\hline

\end{tabular}
\vspace{-10pt}
\end{center}
\end{table}

\subsection{Cross-Checks of \Znunu and \ttbar Background Contributions}
\label{sec:cross_checks}

While the above background estimate is the one used in this search, we
perform auxiliary measurements to cross-check the \Znunu and \ttbar
background  components, which together are expected to comprise the
majority of the background.
As would be crucial in case of an observed excess, these cross-checks
provide an overestimate of the \Znunu and \ttbar background components.

\begin{figure}[tb]
  \centering
  \includegraphics[scale=.36,angle=0]{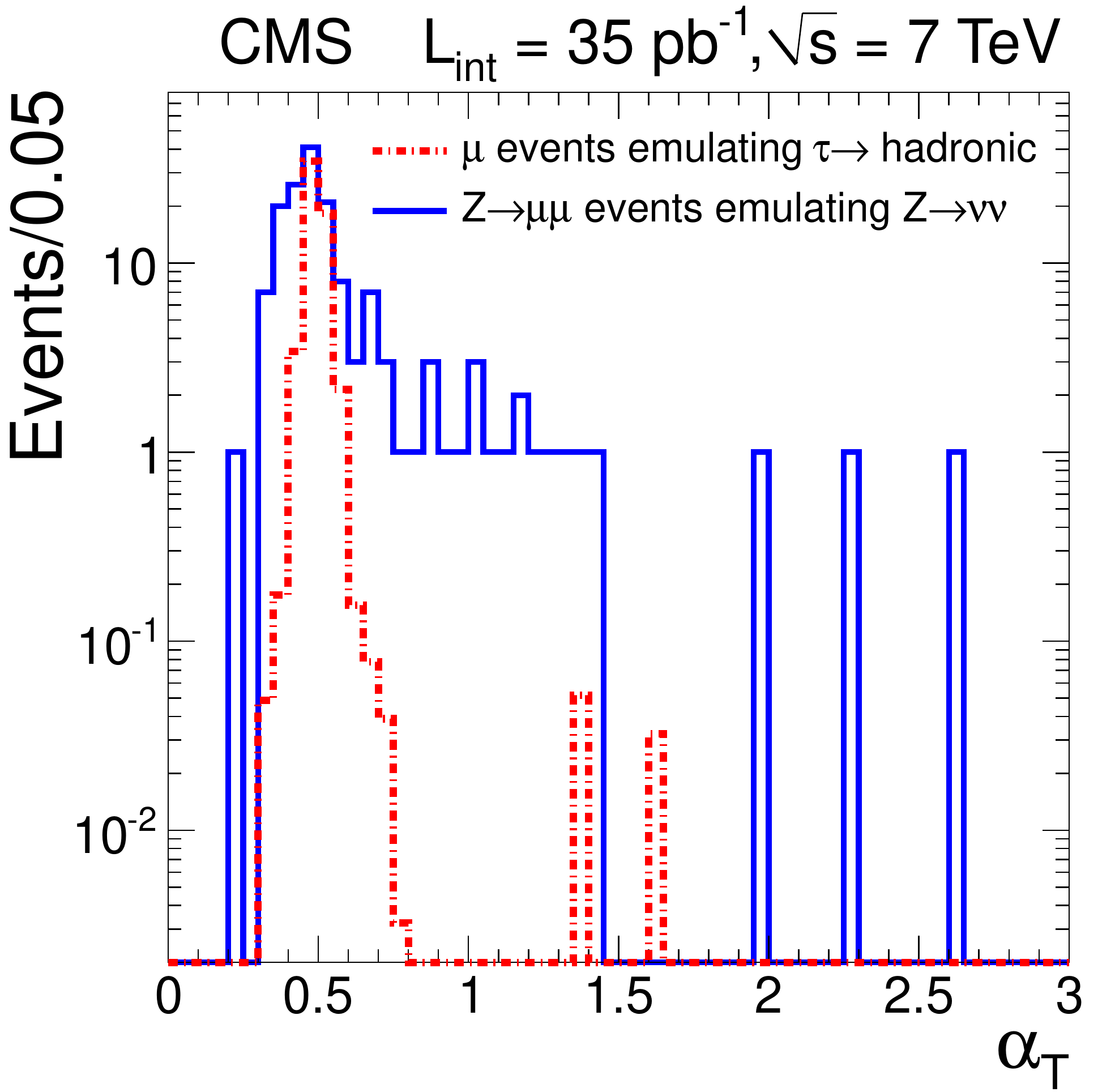}

\caption[\alphat in Zmumu sample with loose jet \et]
  {\label{fig:Zmumu_ttbar_alphat}
   The \alphat distributions for \Zmumu emulation of \Znunu (solid blue) and
   muon emulation of hadronic tau decays (dashed red).
}
\end{figure}

For \Znunu, a sample of \Zmumu events is selected with two or more
jets but no \alphat, \HT, or b-tagging requirements.    The solid blue line in
Figure~\ref{fig:Zmumu_ttbar_alphat} shows the
\alphat distribution for the resulting events.
The fraction of these events
containing a b-tagged jet is measured.
Then, a sample is selected with no b-tag requirement, jet $\et=75$~GeV
thresholds on the two highest-\et jets, \HT\gt275~GeV, and \alphat\gt0.52.
The number of events in this sample is scaled by the measured b-tag
fraction in the other sample, corrected for the muon
identification efficiency and acceptance, and multiplied by the ratio
of branching fractions
$\frac{\textrm{BR}(\Znunu)}{\textrm{BR}(\Zmumu)} \approx 6$.
This procedure gives an overestimate of the number of \Znunu events in the
signal region owing to less stringent requirements than in the
final selection, and yields $0.48\pm0.39$ events.

Simulation studies indicate that most
of the \ttbar background comes from
events with hadronic tau decays.
To estimate the hadronic tau decay yield, \FaT is first measured in a
sample with $\et=80$~GeV thresholds on the two leading jets,
\HT\gt280~GeV, at least one medium b-tagged jet (TCHP~\gt1.91), and one
or two muons.  These selection requirements are chosen to be less
strict than the signal selection in order to increase the number of
events in this sample.
The muons are used to emulate the hadronic decays of taus.
To do so, for each muon the presence of a tau jet is emulated with
an \et value set to a fraction of the muon \pt, using a distribution
taken from simulation.  The dashed red line in
Figure~\ref{fig:Zmumu_ttbar_alphat} displays the resulting
\alphat distribution.
The measured value of \FaT in this sample is multiplied by the number
of emulated events in the signal region before the \alphat
requirement.  This value is corrected for the
muon selection efficiency, acceptance and the hadronic tau decay
branching ratio to obtain the hadronic tau decay yield.
The predicted hadronic tau decay yield is increased by 38\%,
as determined in simulation, in order to account for the entire \ttbar
background.
The procedure yields a 25\% overestimate of the total \ttbar
background in simulation.  In data, $1.4\pm0.5$ events are predicted.

\section{Signal Selection Efficiency}
\label{sec:sig_acc_eff}

To interpret the results of this search in terms of a given signal model,
the selection efficiency for that model must be determined.
Table~\ref{table:acceptance} lists
the cumulative and individual efficiencies for the event selection
in the three SUSY benchmark models LM0, LM1, and LMB, from which
events are generated at leading order (LO) via \PYTHIA~6.4, tune
Z2~\cite{pythia6} using parton distribution functions provided by
CTEQ6.6~\cite{cteq}.
Without b tagging, the cumulative efficiencies for LM0 and LM1
are about 85\% of those in Ref.~\cite{main_RA1_note}, because of a
more stringent lepton and photon veto.
Table~\ref{table:systematics} lists the relative systematic uncertainties on
the signal yield, which are  dominated by the
uncertainty on the b-tagging efficiency,
described below.
The other uncertainties and the
methods used to obtain them are similar to
Ref.~\cite{main_RA1_note}.

\begin{table}[tb]
\caption[Fit Parameters] {\label{table:acceptance}
Cumulative and individual efficiencies for the selection
in three SUSY benchmark points.
For each point, the left and right columns represent the individual
and cumulative efficiencies, respectively.
Different benchmarks have
different b-tag efficiencies due to
different average numbers of b quarks per event.
The fraction of events containing at least one b quark before the
b-tag selection is 66\% in LM0, 18\% in LM1, and 91\% in LMB.
}
\begin{center}
\begin{tabular} { | l | c c | c c | c c |}
\hline

\multicolumn{1}{ | c | } {   } & \multicolumn{2}{  c | } {  LM0 } &
\multicolumn{2}{  c | } {  LM1 }  & \multicolumn{2}{  c | } {  LMB }  \\

Requirement & & & & &  &  \\

\hline

Pre-selection  & $98\%$ & $98\%$ & $98\%$ & $98\%$ & $98\%$ & $98\%$  \\
Lepton/Photon Veto & $57\%$ & $56\%$ & $55\%$ & $54\%$ & $61\%$ & $60\%$  \\
Jet Requirements & $51\%$ & $28\%$ & $63\%$ &
$34\%$ & $54\%$ & $33\%$ \\
\HT\gt350~GeV & $90\%$ & $25\%$ & $94\%$ & $32\%$ & $97\%$ & $32\%$ \\
Trigger & $99\%$ & $25\%$ & $99\%$ & $32\%$ & $99\%$ & $31\%$ \\
Apparent \mht Veto & $68\%$ & $17\%$ & $81\%$ & $26\%$ & $65\%$ & $20\%$ \\
Tight b-tag  & $31\%$ & $5.3\%$ & $12\%$ & $3.0\%$ & $54\%$ & $11\%$  \\
\alphat\gt0.55 & $14\%$ & $0.7\%$ & $29\%$ & $0.9\%$ & $12\%$ & $1.3\%$  \\

\hline

\end{tabular}
\vspace{-10pt}
\end{center}
\end{table}

\begin{table}[tb]
\caption[Fit Parameters] {\label{table:systematics}
Relative systematic uncertainties on the signal yield.
}
\begin{center}
\begin{tabular} { | l r |}
\hline
Source & Uncertainty ($\%$) \\
\hline

Luminosity & $4$ \\
JES        & $3.5$ \\
Jet Energy Resolution & $1$ \\
Trigger Efficiency & 1  \\
Apparent \mht Veto & $4$  \\
Lepton/Photon Veto & $4$  \\
b-tag Efficiency (LMB) & $20$ \\
\hline
Total     & $22$ \\
\hline

\end{tabular}
\vspace{-10pt}
\end{center}
\end{table}

The b-tagging efficiency is measured from inclusive dijet events in
which one jet has an associated muon and another ``away'' jet has a TCHP
value of at least 1.0.
The relative fraction of jets from b quarks in a data sample is
determined by a fit to the distribution of transverse momentum of
muons relative to their
associated jet axis, \ptrel~\cite{BTV10-001,ptrel}, which is larger for jets
from  b quarks than from other flavours.  This fit is to a linear
combination of simulation-derived \ptrel expected distributions from
different flavours.
The fitted b fractions for jets passing and failing the analysis
b-tagging requirement are used in the b-tagging
efficiency calculation.  This efficiency is measured
separately for jets with $|\eta| > 1.4$ and $|\eta| \leq 1.4$, in four
ranges of jet \et.  The ratio between the b-tagging efficiency
measured in data and in simulation is taken as the
efficiency scale factor for a particular range in \et and $|\eta|$.

Systematic uncertainties on the scale factors arise from potential
biases in the \ptrel fitting procedure.  These uncertainties are measured by varying
the muon-to-jet matching and muon \pt thresholds, fraction of gluon splitting
to ${\rm b\overline{b}}$, jet energy scale and resolution, jet angular resolution,
and b-tagging requirement on the away jet.
The effect of measuring the scale factors using only
semi-leptonic b decays is also accounted for.
The scale factors are used to correct the expected event yield at
each signal point for differences between the efficiencies in data and
simulation.  For example, for LMB the application of the scale
factors translates into a change in the yield by a factor
$0.87\pm0.18$.
The systematic and statistical uncertainties give a
total relative uncertainty of 20\% in LMB, with a similar uncertainty of 23\% in LM1.

\section{Results}
\label{sec:results}

The observation of one data event in the signal region is
consistent with background expectations.
Combining the expected signal and background prediction from
Section~\ref{sec:alphaT-vs-HT} and using
frequentist statistical
methods in the manner of Ref.~\cite{Feldman-Cousins} with the
Profile Likelihood ratio~\cite{likelihood} to handle nuisance
parameters, we derive $95\%$ confidence level (CL) cross-section upper limits
(\sigmaobs) of 18.9, 15.4, and 10.2~pb for LM0, LM1, and LMB,
respectively.  The effect of possibly overestimating the background
due to signal contamination in the control regions increases the
\sigmaobs value to 22.1~pb for LM0, 16.7~pb for LM1, but is negligible
for LMB.
To quantify the sensitivity with reduced dependence on the amount of
b-quark production,
a 95\% CL upper limit on the
cross section times branching ratio to at least one b quark of
4.0~pb is determined in LM1.

\begin{figure}[tb]
  \centering
  \includegraphics[scale=.6]{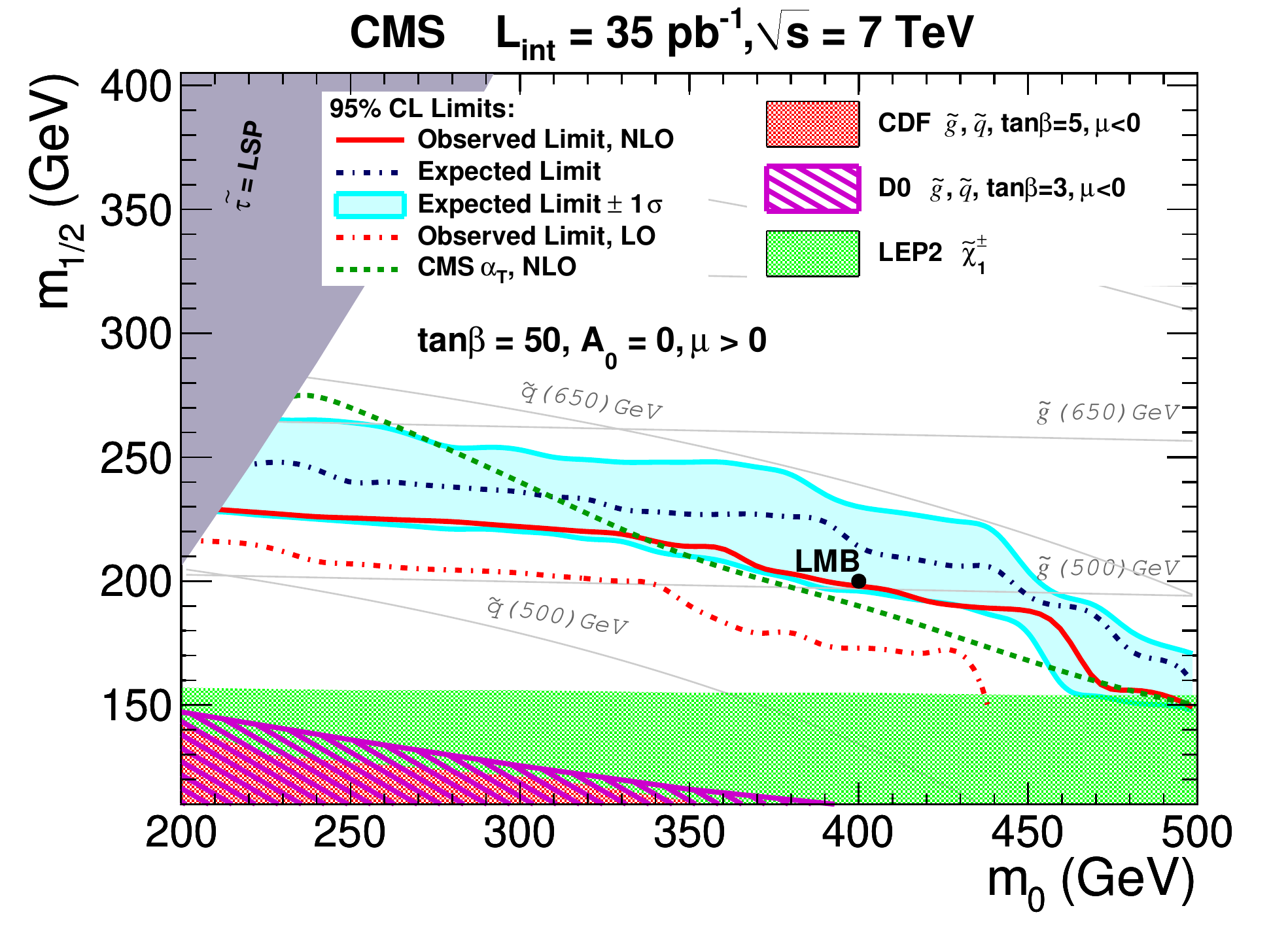}
\caption[CMSSM tanbeta=50 exclusion region]
  {\label{fig:exclude}
    Exclusion regions in the ($m_{0}$, $m_{1/2}$) plane for one set
  of CMSSM parameters, for this analysis (solid red), and the non-b
  tagged version~\cite{main_RA1_note} (dashed green).
}
\end{figure}

The resulting excluded region in the  ($m_{0}$, $m_{1/2}$)  plane for
a reference model with CMSSM parameters $A_{0}=0$~GeV,
  $\tan{\beta}=50$, and $\mu>0$ is shown in Figure~\ref{fig:exclude}.
The expected and observed exclusion regions are calculated using
next-to-leading-order (NLO) cross sections, obtained with the program
Prospino~\cite{prospino}.
The excluded region is extended with respect to that of
Ref.~\cite{main_RA1_note} without b tagging, also shown,
for scenarios with increased b production, such as those with $m_{0}$
above 350 GeV.

\section{Summary}
\label{sec:conclusion}

A search for events with multiple jets, at least one of which is
b tagged, and significant transverse momentum imbalance has been
presented.
One event is observed, which is consistent with background expectations.
The dominant background comes from \ttbar production.
The results of
the search are characterized as an exclusion region in CMSSM parameter
space and 95\% CL upper limits on representative scenarios with
expected cross section of approximately 15~pb.
The sensitivity of this search surpasses that of the Tevatron
experiments~\cite{CDFsearch,D0search}
and is comparable to a recent
fully hadronic, b-tagged search from
ATLAS~\cite{Atlas_allhadronic_btag}.

\section*{Acknowledgements}
\label{sec:acknowledgements}

We wish to congratulate our colleagues in the CERN accelerator departments
for the excellent performance of the LHC machine. We thank the technical
and administrative staff at CERN and other CMS institutes, and acknowledge
support from: FMSR (Austria); FNRS and FWO (Belgium); CNPq, CAPES, FAPERJ,
and FAPESP (Brazil); MES (Bulgaria); CERN; CAS, MoST, and NSFC (China);
COLCIENCIAS (Colombia); MSES (Croatia); RPF (Cyprus); Academy of Sciences
and NICPB (Estonia); Academy of Finland, MEC, and HIP (Finland); CEA and
CNRS/IN2P3 (France); BMBF, DFG, and HGF (Germany); GSRT (Greece); OTKA and
NKTH (Hungary); DAE and DST (India); IPM (Iran); SFI (Ireland); INFN
(Italy); NRF and WCU (Korea); LAS (Lithuania); CINVESTAV, CONACYT, SEP,
and UASLP-FAI (Mexico); MSI (New Zealand); PAEC (Pakistan); SCSR (Poland);
FCT (Portugal); JINR (Armenia, Belarus, Georgia, Ukraine, Uzbekistan);
MST, MAE and RFBR (Russia); MSTD (Serbia); MICINN and CPAN (Spain); Swiss
Funding Agencies (Switzerland); NSC (Taipei); TUBITAK and TAEK (Turkey);
STFC (United Kingdom); DOE and NSF (USA).

\bibliography{auto_generated}   

\cleardoublepage\appendix\section{The CMS Collaboration \label{app:collab}}\begin{sloppypar}\hyphenpenalty=5000\widowpenalty=500\clubpenalty=5000\input{SUS-10-011-authorlist.tex}\end{sloppypar}
\end{document}

%% file: SUS-10-011-authorlist.tex
\textbf{Yerevan Physics Institute,  Yerevan,  Armenia}\\*[0pt]
S.~Chatrchyan, V.~Khachatryan, A.M.~Sirunyan, A.~Tumasyan
\vskip\cmsinstskip
\textbf{Institut f\"{u}r Hochenergiephysik der OeAW,  Wien,  Austria}\\*[0pt]
W.~Adam, T.~Bergauer, M.~Dragicevic, J.~Er\"{o}, C.~Fabjan, M.~Friedl, R.~Fr\"{u}hwirth, V.M.~Ghete, J.~Hammer\cmsAuthorMark{1}, S.~H\"{a}nsel, M.~Hoch, N.~H\"{o}rmann, J.~Hrubec, M.~Jeitler, W.~Kiesenhofer, M.~Krammer, D.~Liko, I.~Mikulec, M.~Pernicka, H.~Rohringer, R.~Sch\"{o}fbeck, J.~Strauss, A.~Taurok, F.~Teischinger, P.~Wagner, W.~Waltenberger, G.~Walzel, E.~Widl, C.-E.~Wulz
% ...
\vskip\cmsinstskip
% ...
1:~~Also at CERN, European Organization for Nuclear Research, Geneva, Switzerland\\
% ...
